# Closed virial equations for hard parallel cubes and squares*

**Leslie V. Woodcock****

**Instituto Ciȩ̂nce Exacte**
**Departamento de Fisica**
**Universidade Federal de Juiz de Fora**
**Brazil**

## SUMMARY

*A correlation between maxima in virial coefficients ($B_n$), and "kissing" numbers for hard hyper-spheres up to dimension D=5, indicates a virial equation and close-packing relationship. Known virial coefficients up to $B_7$, both for hard parallel cubes and squares, indicate that the limiting differences $B_n – B_{(n-1)}$ behave similar to spheres and disks, in the respective expansions relative to maximum close packing. In all cases, the increment $B_n – B_{(n-1)}$ will approach a negative constant with similar functional form in each dimension. This observation enables closed-virial equations-of-state for cubes and squares to be obtained. In both the 3D and 2D cases, the virial pressures begin to deviate from MD thermodynamic pressures at densities well-below crystallization. These results consolidate the general conclusion, from previous papers on spheres and disks, that the Mayer cluster expansion cannot represent the thermodynamic fluid phases up to freezing as commonly assumed in statistical theories.*

## CONTENT

* *Este papel e dedicadoa a memoria de* **Orozina Feliciano de Almeida (1956-2011),** ***de um povo popular advogada de Juiz de Fora; ela paciente determinada inspirada.***

** **e-mail: les.woodcock@manchester.ac.uk**
**permanent address: Manchester Interdisciplinary Biocentre, UK.**

## 1. Introduction

An equation-of-state for the hard-sphere fluid pressure (p) was obtained by analytical closure of the virial expansion in powers of density relative to close-packing [1]. The virial series is simply a Maclaurin expansion of the pressure as a state function about zero density, and can be written in dimensionless form

$$Z = \sum_{n=1}^{\infty} B_n \left( \frac{\rho}{\rho_0} \right)^{n-1} \tag{1}$$

where $Z = pV/Nk_BT$, T is absolute temperature, $k_B$ is Boltzmann's constant, $B_n$ are the dimensionless coefficients $\rho$ is the number density $N/V$, and $\rho_0$ is the crystal close packing density. The theory of the coefficients $B_n$ is well-established; exact statistical expressions for $B_n$ are available as the cluster integrals that are calculated analytically up to $B_4$ and presently available numerically up to $B_{10}$.

A closed form of equation (1) based only upon known coefficients, $B_1$ to $B_{10}$, has been shown to be everywhere convergent up to its first pole at $\rho_0$. The equation-of-state is accurate for the equilibrium hard-sphere fluid, yielding the same thermodynamic pressures as may now be obtained with 6-figure precision from MD simulations of up-to a million spheres [2]. A more refined scrutiny of the margins of difference, with due consideration of the uncertainties, however, shows that the virial equation begins to deviate from the thermodynamic pressure equation, albeit very slightly, at a density in the mid-fluid range close to the available volume percolation transition [3].

It has also been shown that the known virial series of the two-dimensional (D=2) hard-disk fluid is also amenable to closure [4]; an equation-of-state is obtained with 6-figure accuracy at low density. As in the D=3 case, but more pronounced however, there is a clearer deviation well-below the freezing transition. A subsequent more accurate analytical closed form for the pressure equation-of-state for hard disks was derived [5]. The precision of this revised version of the closed virial equation was such that we were able to conclude that the deviation from the thermodynamic pressure occurs at or near the 2D-percolation density of available, or excluded volume, $\rho_{pa}\sigma^2 = 0.4$, determined previously by Hoover et al. [6]. Thus, for both disks (D=2) and spheres (D=3) the virial equations-of-state do not represent the thermodynamic state functions for densities above the respective free volume percolation transitions.

It is also worth noting that the known virial coefficients for D=4 hyper-spheres [7], in powers of density relative to close packing, also increase from $B_1$-$B_6$, then decrease with $B_n$-$B_{(n-1)}$ linearly with n, going negative at n=10. This indicates that the higher virial coefficients should eventually go negative, yielding a closed virial equation with a negative pressure pole at maximum packing in this case also.

These empirical observations regarding the destiny of the virial equations-of-state at higher densities would appear to confirm that the percolation transitions in these hard-core model fluids may indeed be higher-order thermodynamic phase transitions that signify the onset of the divergence of the virial pressure equation from the thermodynamic pressure. There are those who seem to "religiously" believe the Mayer virial series is equivalent to the thermodynamic fluid equation-of-state; this is a common misapprehension amongst physicists.

The suggestion that the virial expansion may not represent the physical thermodynamic state for densities above the percolation transition is not new. For both hard parallel cubes and hard parallel squares, the virial coefficients are known up to $B_7$ [8-10] In fact, it has been known for 50 years that the virial coefficients for parallel cubes go negative [10]. MD calculations 25 years ago [11], of the thermodynamic and transport properties of hard parallel cubes, indicated that the hard-parallel-cube fluid virial equation deviates above the percolation transition, at a density well-below the freezing transition. Kratky [12] further elucidated upon the suggestion the percolation transitions in spheres and disks may be higher order thermodynamic phase transitions. Reference [12] also gives valuable definitions and discussion of the various percolation transitions in hard-core fluids generally.

**Table 1:** Known virial coefficients of the parallel hard-cube fluid (D=3) and parallel hard square fluid (D=2): also given in columns 3 and 5, for D= 3 and 2 respectively, are the predicted values up to $B_{12}$ from the closed virial equations obtained here.

| *n* | $B_n$ [D=3] | Eqn. (3) | $B_n$ [D=2] | Eqn. (6) |
|---|---|---|---|---|
| 1 | 1 | | 1 | |
| 2 | 4 | | 2 | |
| 3 | 9 | | 3 | |
| 4 | 11.33333 | | 3.666666 | |
| 5 | 3.16 | | 3.7222 | |
| 6 | -18.88 | -18.88 | 3.025 | 3.0407 |
| 7 | -43.5 | -43.50 | 1.65 | 1.6714 |
| 8 | | -69.00 | | -0.3857 |
| 9 | | -94.50 | | -3.1306 |
| 10 | | -120.00 | | -6.5633 |
| 11 | | -145.50 | | -10.6838 |
| 12 | | -171.00 | | -15.4921 |

On revisiting the known virial coefficients for both fluid systems of parallel cubes and squares (Table 1), it can be seen that in both cases, now in the expansion of the packing fraction ($y = \rho/\rho_0$) even from just the known coefficients up to $B_7$, the differences $B_n$-$B_{(n-1)}$ are approaching negative constants with functional forms analogous to the D=3 spheres [1-3] and D=2 disks [4,5] cases respectively. Here we obtain and test the closed virial equations-of-state for the systems of parallel cubes and squares, which by analogy with the closed equations of spheres [1-3] and disks [4,5], and then compare with available MD data from the literature.

## 2. "Kissing" numbers and virial coefficients

The first simple observation we make, about the $B_n$ values in Table 1, is that for both D=3 and D=2 , $B_n$ first increases with n, peaks at n=4 (3D) or 5 (2D) and then begin to decrease, and in the case of the cubes go negative. This is just the same behavior that is seen generally in hyper-spheres of D > 1, i.e. D=2 , 3 and 4 and 5

The highest virial coefficient in all these cases is of the same order as a maximum co-ordination or contact numbers, known as "kissing numbers" to mathematicians [12], of contiguous spheres to a central sphere. This correlation provides salient evidence that the virial expansions about zero density, in powers of density relative to close packing, equation (1), reflects information about the close packed state, and is *per se* suggestive of the closed forms with the first poles at that maximum density.

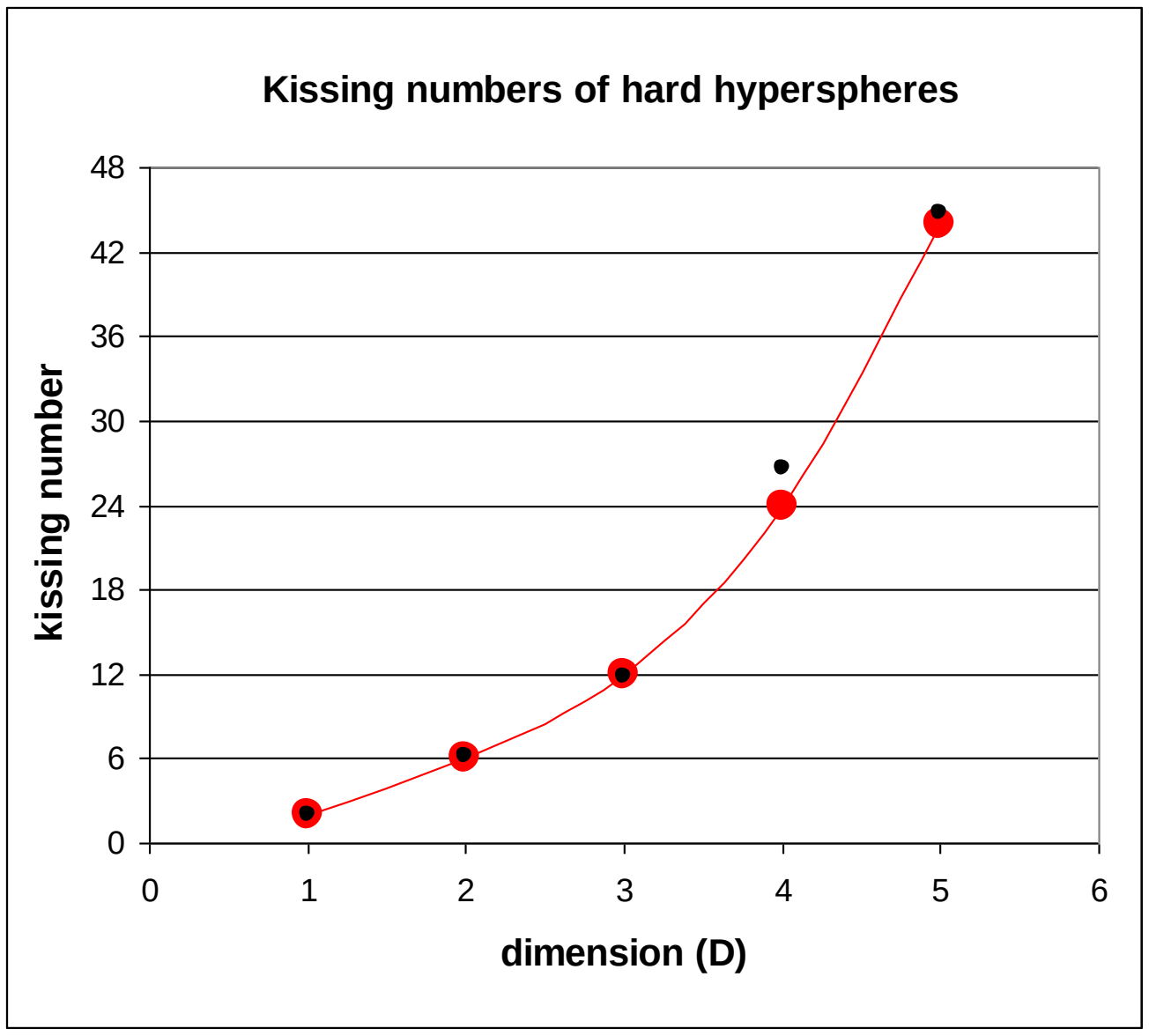

**Figure 1**: Kissing numbers for D-dimensional hyper-spheres (shown as red circles centered on the exact whole number) compared to the values of $B_n$(max) + D (shown as smaller black circles); the maximum virial coefficients for D = 2 to 5 are 4.243 ($B_9$), 8.864($B_8$), 22.639($B_6$) and 39.784($B_5$) respectively. The kissing number is simply defined as the maximum number of hyper-spheres in a configuration of like spheres that may be contiguous with a central sphere at the same time. This number is not necessarily the lattice near-neighbor number, but for the lower dimensions up to D=5, it can represent the first co-ordination number in the most stable crystal structure at infinite pressure, or equivalently, at zero temperature. Kissing numbers for hyper-spheres are known up to D=28 [12] but here (Figure 1) we compare with D=1 to D=5. The correlation is convincing evidence that the virial expansion of non-space-filling hyper-spheres does not reflect the physically unreal packing (y=1), seen in many theoretical approximate virial equations, but rather the maximum crystal close-packed states ($\rho_0$).

Looking again at Table 1, the $B_n$ values for the cases D= 3 (cubes) and D=2 (squares) have similar maxima at $B_4$ (D=3) value 11.3333 and $B_5$ (D=2) value 3.7222. There are three different types of "kisses" in these systems, faces, edges and corners. Counting only faces and edges, we get for $B_{max}$ +D, 14.3 and 5.7, compared to 14 and 4 for the respective kissing numbers of cubes and squares respectively. In these cases, the density relative to close packing $\rho/\rho_0$ = packing fraction y, and the maximum density corresponds to y=1.

Below, we will use y to represent reduced number density, and anticipate a singularity in the closed virial equations at y=1.

## 3. Hard parallel cubes

The values of the known virial coefficients for hard parallel cubes are listed in Table 1. Although presently limited to $B_7$, the values $B_5$ to $B_7$ are already indicative of a similar closed functional form to that obtained for hard-spheres [1,2]. In Figure 2, when plotted against exp(-n), by analogy with spheres in reference [2], the incremental values $B_n$-$B_{(n-1)}$ beyond ($B_5$-$B_4$) decrease exponentially as $n \rightarrow \infty$ according to

$$B_n - B_{(n-1)} = A_0 + A_1 \exp(-n) + A_2 \exp(-2n) \tag{3}$$

and approach a constant value (25.5) which $B_7$-$B_6$ (= 24.62) is almost upon.

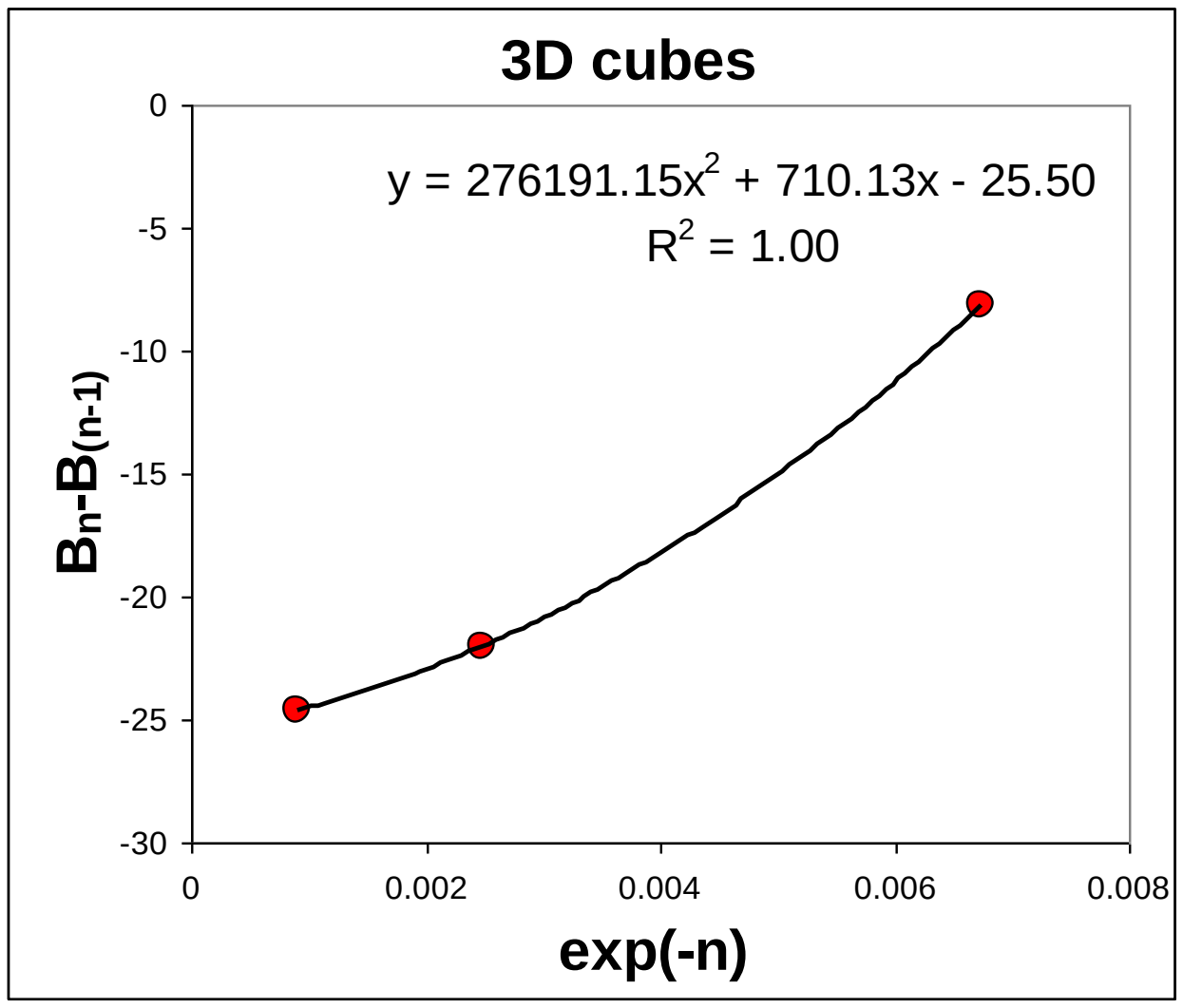

**Figure 2:** Difference between successive coefficients ($B_n$) from n= 7 to n = 5 in the expansion in powers of the density relative to close packing: the difference $B_n$- $B_{(n-1)}$ decreases exponentially , and approaches the constant -25.5 when $n \rightarrow \infty$ .

Neglecting the higher order term, once $A_0$ is obtained, the constant $A_1$ is determined from $B_m$-$B_{m-1}$ and can be used to predict all higher values $B_n$ from m+1 to infinity, thereby effecting an analytic closure to the virial expansion.

As with hard spheres, interpolation of the known virial coefficients suggests that since the limiting value of $B_n$-$B_{n-1}$ is negative, the virial coefficients beyond $B_7$ will remain negative. The corresponding virial equation-of-state will be continuous in all its derivatives, eventually showing a negative pressure, with the first pole at $\rho_o$, i.e. y=1.
The virial equation-of-state for hard parallel cubes would then take the form

$$Z = 1 + \sum_{n=2}^{m} B_n y^{(n-1)} + \sum_{n=m+1}^{\infty} (B_m + A_0 + A_1 e^{-(m+1)}) y^{(n-1)} \tag{4}$$

In which each term of the second summation can be closed separately as with spheres, [2,3] to obtain an equation-of-state. Equation (4) enables the closure for any known n greater or equal to m=7. The last term is a negligible correction for finite m and disappears for large m. It would appear from the data in Figure 1 and Table 1 that m=7 is sufficiently large for accuracy of the same order as that of the available MD pressures in this case, which is about 4-figure precision. Accordingly, if we close the summation at m= 7 and put $A_o$ = - 25.5, equation (4) reduces to a form which is the same as the closed virial equation for hard spheres [1,2] with $A_0$ for cubes fixed at -25.5 (Figure 2).

$$Z = 1 + \sum_{n=2}^{m} B_n y^{(n-1)} + B_m \frac{y^m}{(1-y)} - A_0 \frac{y^m}{(1-y)^2} \tag{5}$$

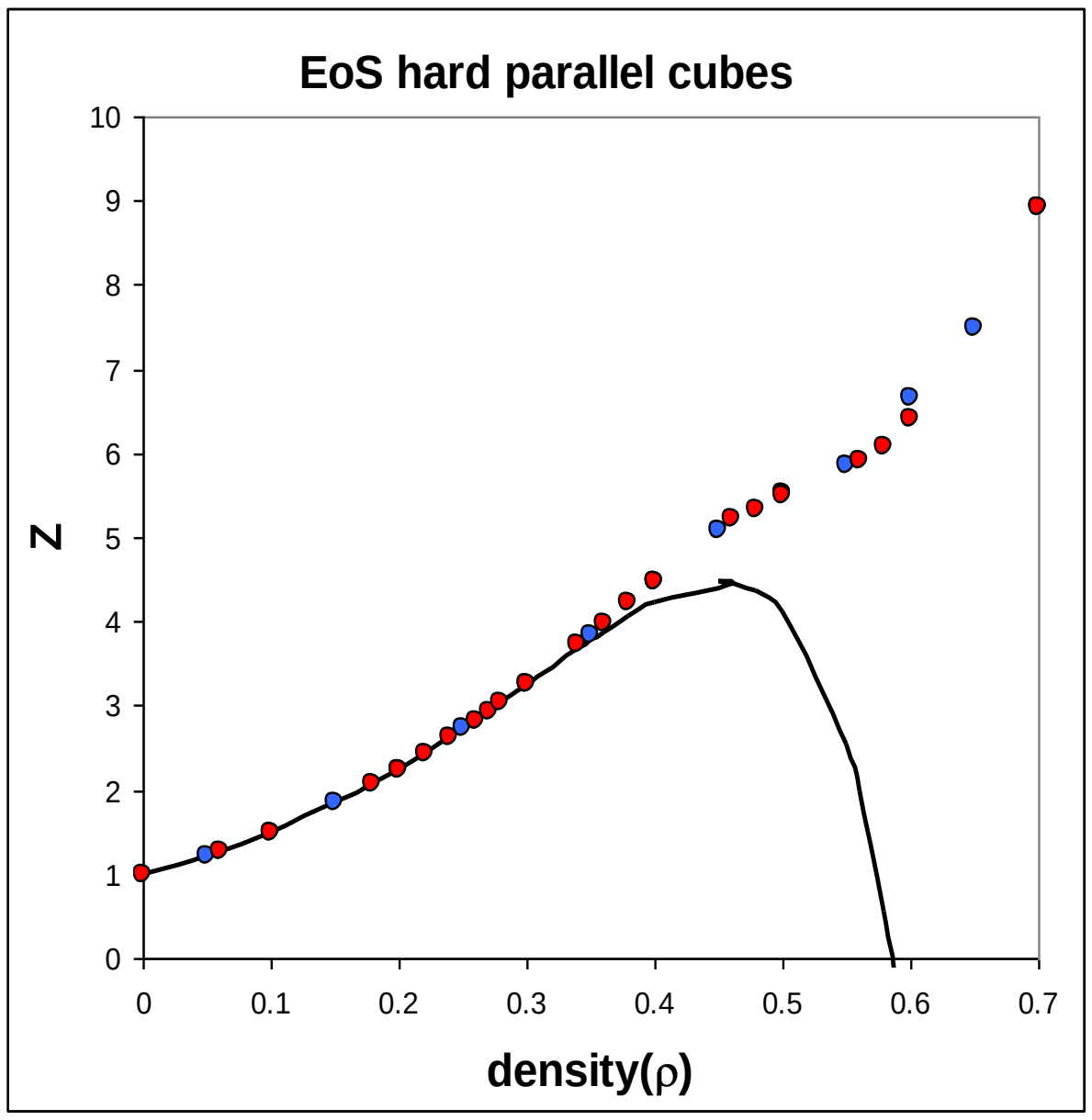

**Figure 3**: Closed-virial equation-of-state (equation (4) with *m*= 7 and $A_0$ = -25.5 : solid line) compared with thermodynamic pressures ($Z = pV/Nk_BT$) obtained from MD simulations by Woodcock and van Swol [11] (red circles) and Hoover et al. [14] ( blue circles).

This equation-of-state for parallel cubes (equation (5) with $A_0$= -25.5) which can now be compared with the literature values of available thermodynamic pressures obtained from MD simulations [11,14]. The virial equation it accurate to within the uncertainty in the experimental data up to the density around 0.2 to 0.3 and then it begins to deviate. The pressure peaks, then decreases and goes negative, and eventually diverges with a negative pole at the density of maximum packing. The difference in pressure between the closed virial equation (5), with A0= -25.5, and the thermodynamic pressure from MD computations in the vicinity of melting is shown in Figure 4.

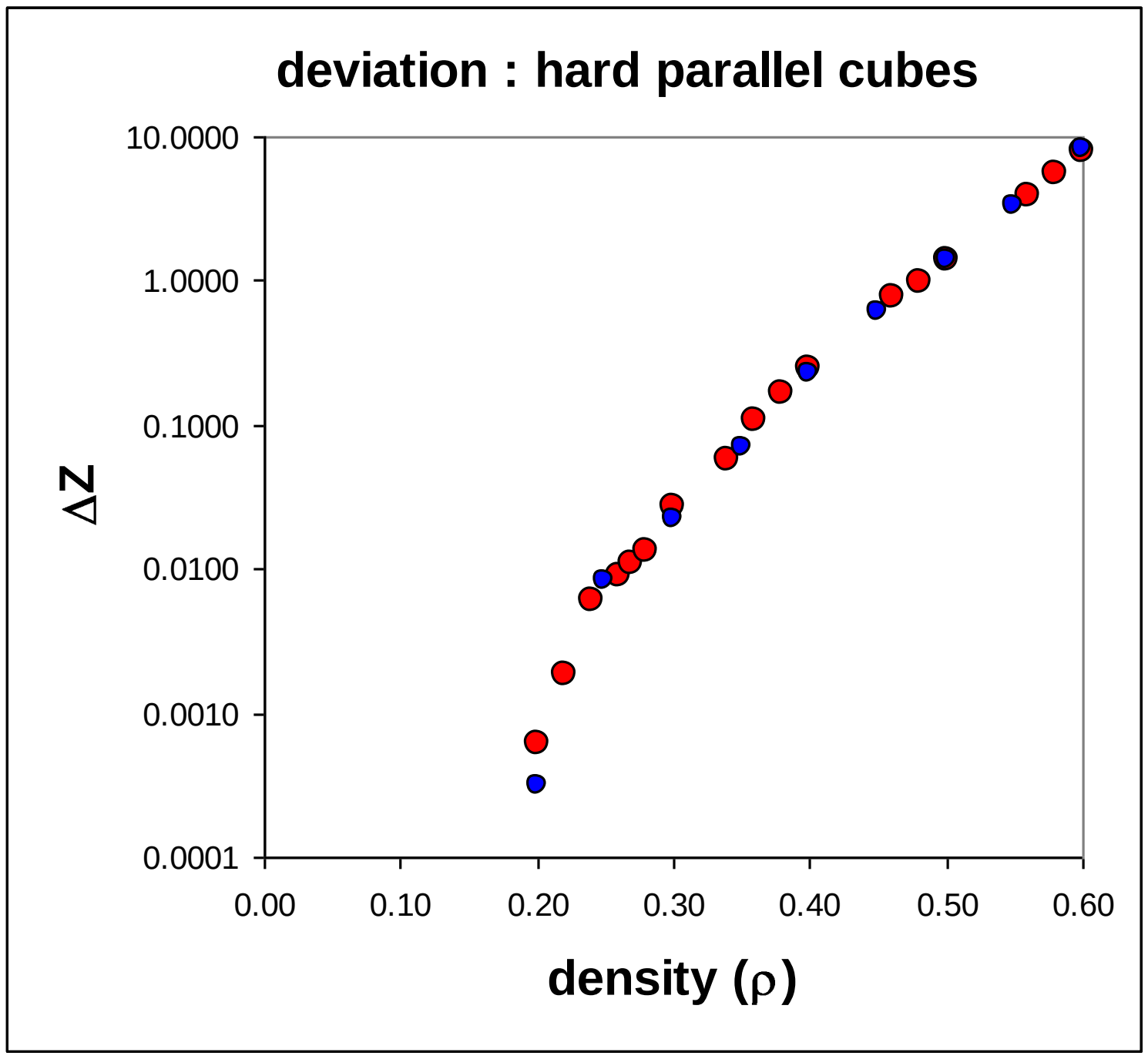

**Figure 4**: Deviation of closed-virial equation-of-state (equation (5): m=7, $A_0$=25.5) and thermodynamic pressures obtained from MD simulations by Woodcock and van Swol [11](red circles) and Hoover et al. [13] (blue circles): $\Delta Z = Z_{MD} - Z_{(virial)}$

Inspection of the discrepancy, between closed virial equation (5), and thermodynamic MD pressures, in Figure 4, suggests that the virial pressure begins to deviate from the thermodynamic pressure at a density on or below the available volume percolation density ($y_{pa}$) [11]. This deviation is statistically significant; it would appear from the forgoing analyses that it cannot be explained either by uncertainties in the thermodynamic pressures, or errors in the known virial coefficients, or any combination of both. To do so the virial coefficients $B_8$ and beyond would have to take an extraordinary unusual twist. As with spheres, however, we cannot say that within the uncertainties, the deviation does not begin at an even lower density, perhaps at the lower percolation transition density ($y_{pe}$) associated with the excluded volume (see also references [3,12]).

## 4. Hard parallel squares

Turning now to the 2-D case of hard parallel squares, inspection of incremental values of successive virial coefficients in Table 1 shows that squares are behaving similar to 2-D disks. Differences in successive $B_n$, in powers of density relative to crystal close packing in equation (1), $B_{(n)}$-$B_{(n-1)}$ are plotted in Figure 5. With only the three points available the graph that beyond ($B_5$-$B_4$) the increment decrease varies according to

$$B_n - B_{(n-1)} = \alpha_0 + \alpha_1/n \qquad (6)$$

Equation (6) is the same functional form that was obtained for hard disks where the virial coefficients are known up to $B_{10}$. As with disks, this interpolation of the known virial coefficients suggest that since the limiting constant $\alpha_0$ is negative, the virial coefficients will eventually become negative and the corresponding virial equation-of-state will eventually give a negative pressure, with the first pole at y=1, as with cubes. Equation (6) with the parameters obtained from the EXCEL trendline in Figure 5 predicts the first negative coefficient for squares will be $B_8$. In the corresponding closed-virial equation for D=2 disks $B_{31}$ is predicted to be the first negative coefficient [5].

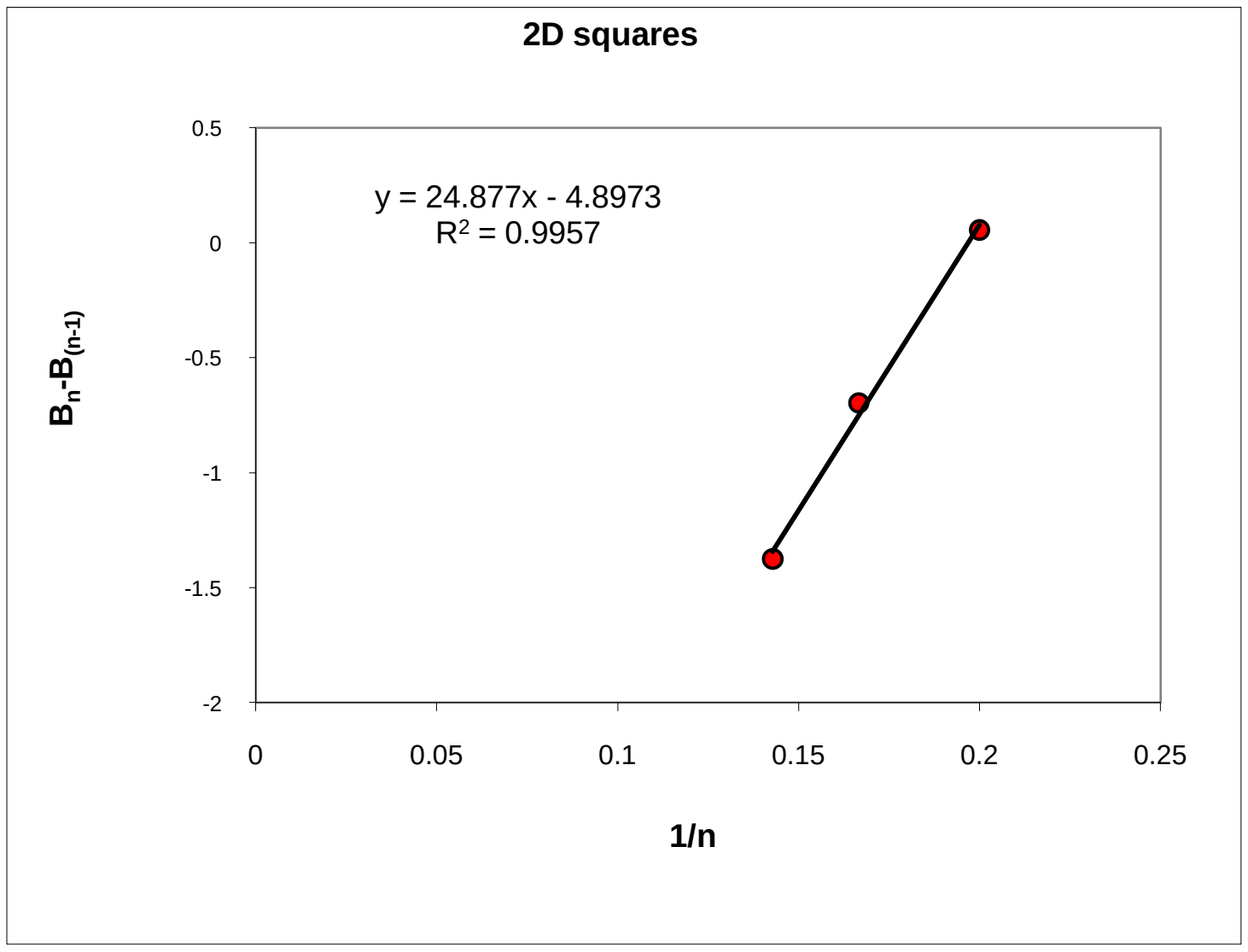

**Figure 5:** Difference between successive coefficients ($B_n$) from n= 7 to n = 5 in the expansion in powers of the density relative to close packing: the difference $B_n$- $B_{(n-1)}$ is decreasing roughly as 1/n, and would approach the constant -4.8973.

By analogy with disks, then the analytic closed form virial equation-of-state for the hard parallel square fluid takes the same form as that of the hard-disk fluid as follows (the derivation is given in the APPENDIX of reference [5]).

$$Z = \sum_{n=1}^{m}\left(B_n - \frac{\alpha}{n(1-y)}\right)y^{(n-1)} + \left(\frac{B_m}{(1-y)} - \frac{\alpha_o}{(1-y)^2}\right)y^m + \frac{\alpha}{y(1-y)}\ln(1-y) \qquad (7)$$

where $B_m$ is the highest known virial coefficient, presently $B_7$.

Using only the known coefficients $B_5$ to $B_7$ as given in Table 1, the constants in the closed virial equation for parallel squares, from the limiting value of $B_n$- $B_{(n-1)}$ Figure 5, are $\alpha_0$ = -4.90, and the slope $\alpha$ = 24.9. The closed virial equation-of-state for the pressure of the parallel-square fluid can then be compared with the thermodynamic pressure from the recent MD computations of Hoover et al. [14]. The virial pressure deviates from the thermodynamic pressure at a density well-below the freezing transition (around y=0.7) eventually gives an unrealistic negative pressure and diverges with a negative pole at maximum close packed density $\rho_0$ corresponding to y=1.

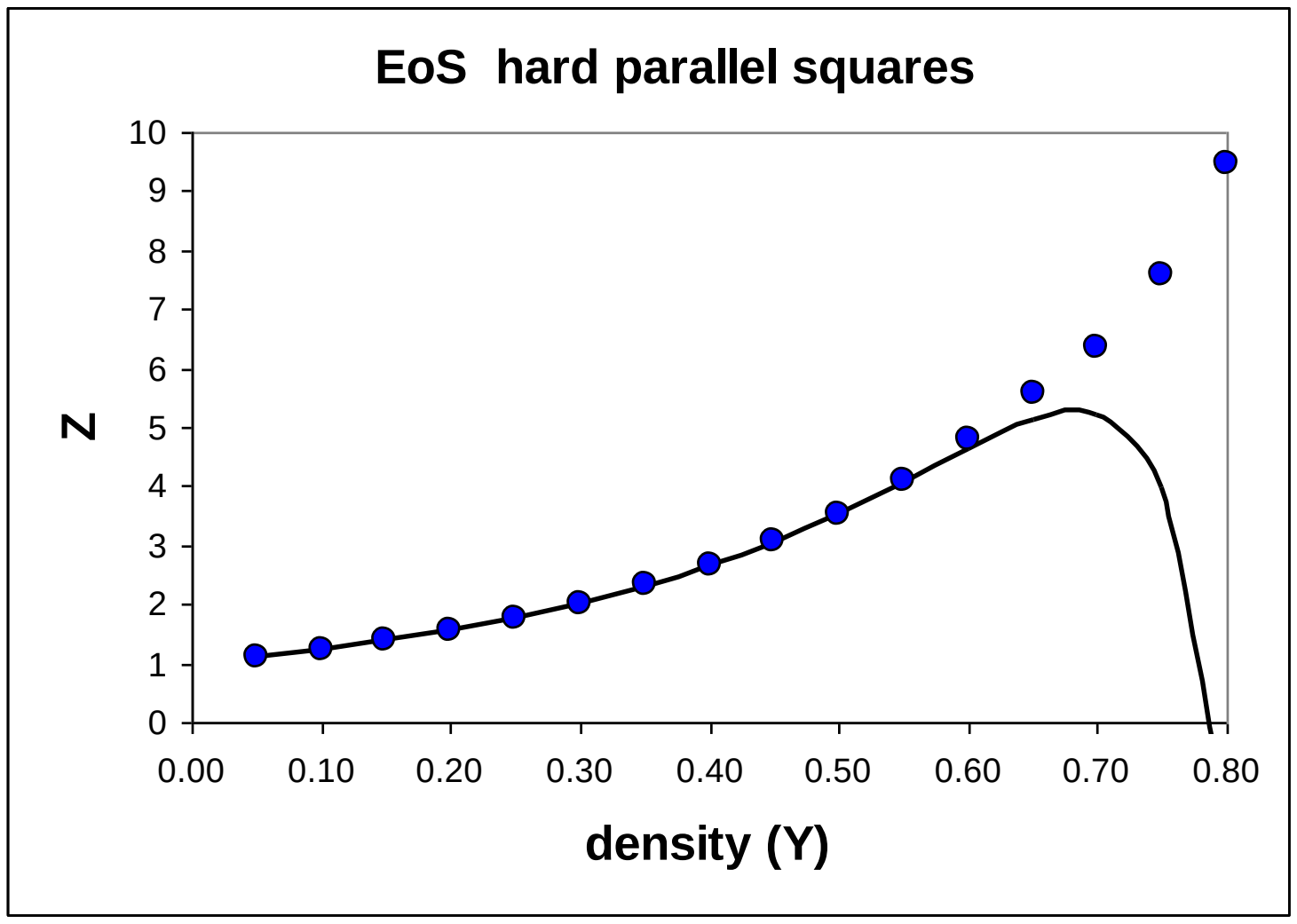

**Figure 6:** Closed-virial equation-of-state (equation (4) with *m*= 7 and the parameters A and $A_0$ as given in Figure 5: solid line) compared with thermodynamic pressures (z = $pV/Nk_BT$) obtained from MD simulations by Hoover et al. [14] (blue circles).

In Figure 7 , the deviation of equation (7) for squares from MD pressures [14] are plotted as a function of density for all the MD data points above in Figure 3 except the data point at the density 0.35 which is an aberration that may contain an error. This plot suggests that the deviation is originating in the vicinity of a low density percolation transition. We have not seen a report of the determination of this percolation transition for squares, but the deviation is around the same value of $\rho/\rho_0$ (= y for squares) as that found by Hoover *et al.* to be the onset of the free volume percolation transition for the D=2 hard-disk fluid [6].

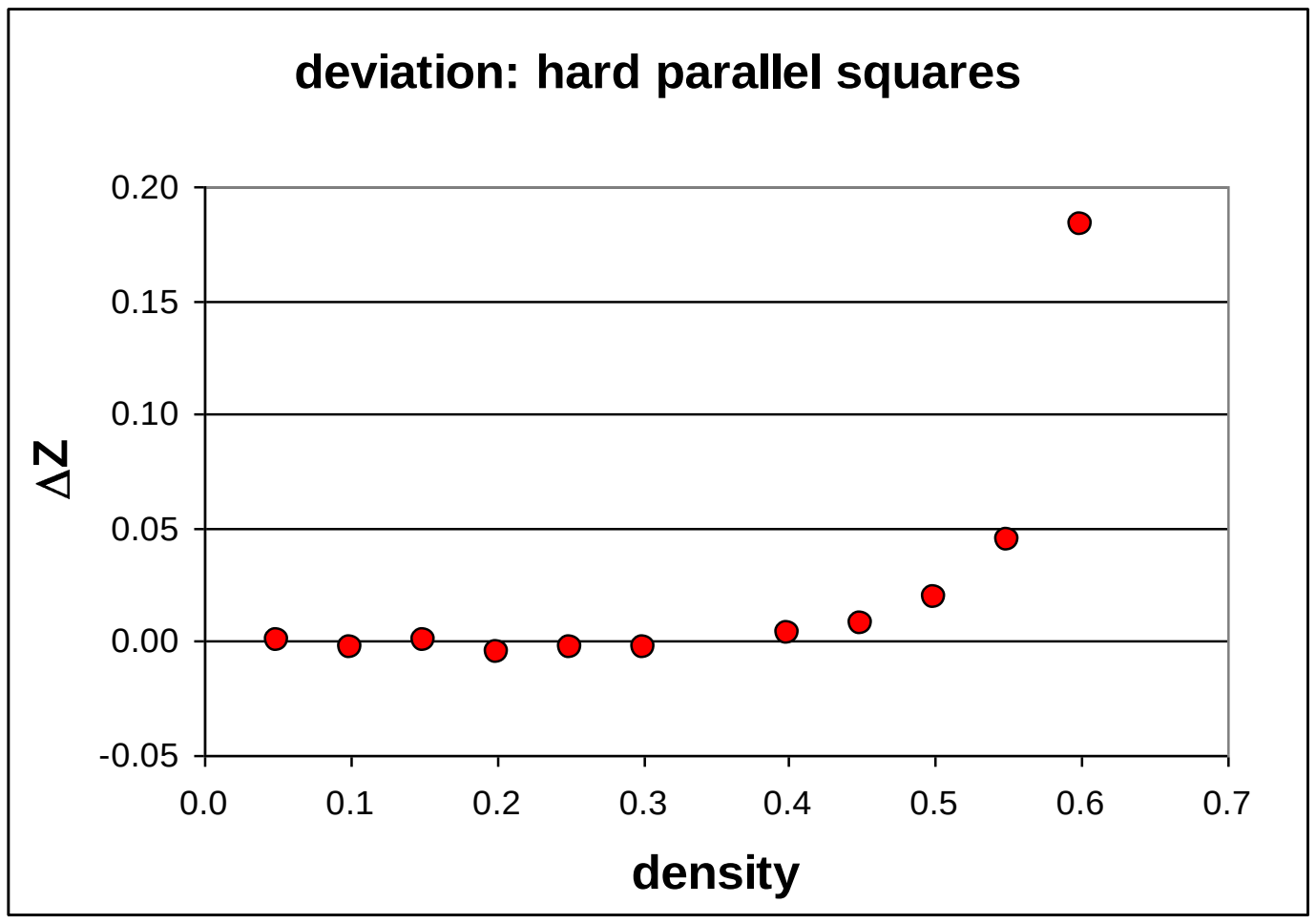

**Figure 7**: Density dependence of pressure difference between closed-virial equation-of-state (equation (4): *m*=10) and thermodynamic pressures obtained from MD simulations by Hoover et al. [14].

## 5. Conclusions

In this paper we have looked at the trend in $B_n$-$B_{n-1}$ for the known virial coefficients of cubes and squares in the Mayer virial expansion equation (1) and observed that the same closed virial equations exhibit the same functional forms, as has been obtained for spheres and disks respectively [1-5].

When the resultant closed vrial equations are compared with available thermodynamic pressures, as with the fluids of spheres and disks, the virial pressure begins deviating from the thermodynamic pressure at a low fluid density. The percolation transition associated with free volume has only been estimated for cubes, and has not yet been reported for squares. Nonetheless, all the evidence is that the onset of the deviations may be associated with higher-order thermodynamic phase transitions.

The APPENDIX to this paper illustrates the belief that the virial expansion of Mayer [15] is actually equivalent to the fluid equation-of-state of these hard core models, at least up to freezing, is a widespread misapprehension amongst theoretical physicists. In response to the various suggestions that the empirical results of these closed-virial comparisons are "speculative", it seems that what has been unduly "speculative" is the incorrect assumption that Mayer's cluster expansion represents the thermodynamic state functions of the fluid phases up to or beyond the freezing transition. In five cases we have so far looked at, first spheres and disks , now here squares and cubes, and also D=4 hyper-spheres (unpublished), the virial equations are failing at a low equilibrium-fluid density.

Many statistical theories are based upon the belief that, if all the terms in the Mayer cluster expansion could be approximated accurately, it would be a theory of “liquids”. We now see that all these simple hard-core models have two fluid phases, the low density gas phase where the Mayer virial expansion represents the thermodynamic state functions, and the high density fluid phase where it is invalid. Looking again at the theory of simple liquids, we may now conjecture that the high density fluid phase belongs to the same phase as the supercooled “liquid” phase, by definition. For hard-spheres this is the metastable branch that is a continuous extrapolation of the equilibrium high-density fluid at freezing, and which terminates at the random close packed (RCP) state. Perhaps we should now take another look at the RCP state as a starting point for the general theory of liquids.

Standard treatises on simple liquids’ deal largely with theories of the liquid state based upon “configurational surgery” of Mayer virial cluster diagrams [15,16]. We now see that the Mayer cluster expansion whilst being an essentially exact theory of low density gases, may not be a starting point for theories of the liquid state. An analytical theory with all the virial coefficients correct would not still represent the high density equilibrium fluid. The title of Hansen and McDonald [16], when the 4$^{th}$ Edition comes to be published, might be retitled “The Theory of Simple Gases”!

## Acknowledgements

I wish to thank the Department of Physics at the Federal University of Juiz de Fora for a temporary visiting facility, and my host Professors, Socrates Danas and Bernhard Lesche, for their support and kind hospitality.

## References

[1] Woodcock, L.V., “Virial equation-of-state for hard spheres”, LANL Ar. Xiv cond-mat. 0801.4846 [pdf] (2008).

[2] Bannerman M., Lue L. and Woodcock L. V., “Thermodynamic pressures for hard spheres and closed virial equation-of-state”, J. Chem. Phys. 132 084507 (2010).

[3] Woodcock, L. V., “Percolation transitions in the hard-sphere fluid”, AIChE Journal (accepted and published online April 2011).

[4] Woodcock, L.V., “Virial equation-of-state for the hard disk fluid”, 2008 LANL *Ar. Xiv cond-mat*. 0804.0679 [pdf](2008)

[5] Beris, A. and Woodcock L.V. “Closed virial equation-of-state for the hard-disk fluid”, LANL Cond. Mat. Stat. Mech. arXiv:1008.3872 [pdf] (2010).

[6] W. G. Hoover, N. E. Hoover and K. Hanson, "Exact hard-disk free volumes", J. Chem. Phys. 70 (04), 1837-1844 (1979).

[7] N. Clisby and B. N. McCoy, Ninth and tenth virial coefficients for hard hyper-spheres in D-dimensions. J Statistical Physics, 122, 15-57 (2006).

[8] R. W. Zwanzig, "Virial coefficients of parallel square and parallel cube gases", J. Chem. Phys. 24, 855-856 (1956).

[9] W. G. Hoover and A. G. De Rocco, "Sixth virial coefficients for gases of parallel hard lines, squares, and cubes", Journal of Chemical Physics 34, 1059-1060 (1961).

[10] W. G. Hoover and A. G. De Rocco, "Sixth and seventh virial coefficients for the parallel hard cube model", J. Chem. Phys., 36, 3141-3162 (1962).

[11] van Swol F. and Woodcock L. V., Percolation transition in the parallel hard cube model fluid, *Molecular Simulation*; 1 (1), 95-108 (1987)

[12] Kratky, K. W., "Is the percolation transition of hard spheres a thermodynamic phase transition?" J. Statistical Physics. 52 , 1413-1421 (1988)

[13] see e.g. www. http://mathworld.wolfram.com/KissingNumber.html

[14] W. G. Hoover, C. Hoover and M. Bannerman, "Single speed molecular dynamics of hard parallel squares and cubes", J. Statistical Physics 136 (4) 715-732 (2009)

[15] Mayer J.E. and Mayer, M.G., "Statistical Mechanics" (John Wiley, New York: 1940)

[16] Hansen J-P. and McDonald I. R., The Theory of Simple Liquids, 1$^{st}$ Edition (1976), 2$^{nd}$ Edition (1989) 3$^{rd}$ Edition (2006) (Academic Press: Oxford)